\documentclass[fleqn,10pt]{wlscirep}
\usepackage[utf8]{inputenc}
\usepackage[T1]{fontenc}
\title{Higgs-like stiffness and fractons on the verge of phase transitions}

\author[1]{Lucas Squillante}
\author[2]{Antonio C. Seridonio}
\author[1]{Roberto E. Lagos-Monaco}
\author[1,*]{Mariano de Souza}
\affil[1]{S\~ao Paulo State University (Unesp), IGCE - Physics Department, Rio Claro - SP, Brazil}
\affil[2]{S\~ao Palo State University (Unesp), Department of Physics and Chemistry, Ilha Solteira - SP, Brazil}

\affil[*]{mariano.souza@unesp.br}

\begin{abstract}
In condensed matter Physics, massive longitudinal Higgs modes emerge from fluctuations of the order parameter.  A few years ago, the Higgs mode was \emph{caught} experimentally in the vicinity of an insulator-to-superconductor quantum phase transition [Nat. Phys. \textbf{11}, 188 (2015)]. Here, we propose, in analogy to the Higgs mode, the concept of Higgs-like stiffness (HLS), which emerges close to both classical and quantum phase transitions as a universal manifestation of matter. We build up a Landau free energy for the dielectric response function to demonstrate that \emph{any} complex physical quantity can be used to infer the presence of the HLS. Our analysis is corroborated by experimental results of the quasi-static dielectric constant for the (TMTTF)$_2$SbF$_6$ Fabre salt. Yet, we discuss the appearance of fractons in connection with the locking of particular molecular rotational degrees of freedom.
\end{abstract}
\begin{document}

\flushbottom
\maketitle
\thispagestyle{empty}

%\noindent Please note: Abbreviations should be introduced at the first mention in the main text – no abbreviations lists. Suggested structure of main text (not enforced) is provided below.

\section*{Introduction}
\begin{figure}[!t]
\centering
\includegraphics[width=0.9\columnwidth]{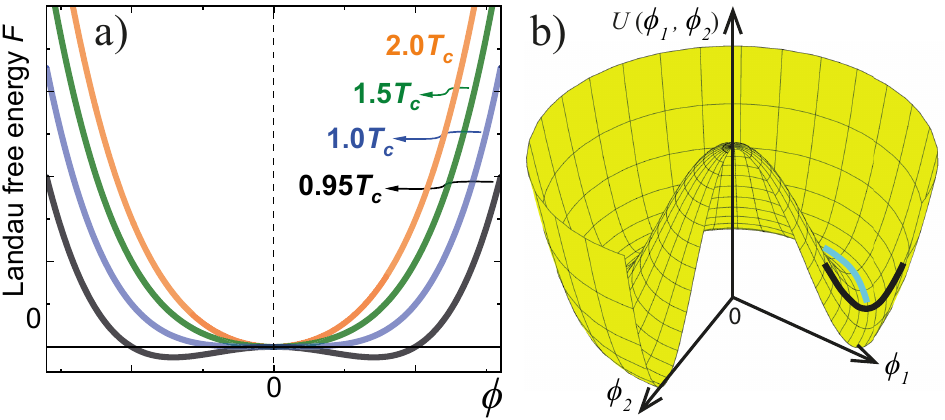}
\caption{\footnotesize a) Landau free energy $F$ \emph{versus} order parameter $\phi$ for various temperatures in terms of the critical temperature $T_c$. b) Potential energy $U (\phi_1, \phi_2)$ \emph{versus} $\phi_1$ \emph{versus} $\phi_2$, where $\phi_1$ and $\phi_2$ are scalar fields. The cyan and black thick solid lines represent, respectively, the Nambu-Goldstone and the Higgs modes. The 3D plot of $U(\phi_1, \phi_2)$ is partially shown for a better visualization of the Nambu-Goldstone and the Higgs modes. More details in the main text.}
\label{Fig-1}
\end{figure}

In his seminal paper about \emph{plasmons, gauge invariance, and mass}, P.W.\,Anderson proposed that ``... \emph{plasma frequency is equivalent to mass} ...'' \cite{Anderson1963}. A key question which arises is how mass can be ``created'' in this framework? It is well-established that a continuous symmetry breaking gives rise to two excitation modes, namely the transverse massless Nambu-Goldstone \cite{nambu, goldstone} and the longitudinal massive Higgs modes \cite{higgs}, usually identified using a potential with a ``Mexican hat'' shape, cf.\,Fig.\,\ref{Fig-1}. Recently, the Higgs mode was experimentally accessed in disordered thin films of NbN and InO superconductors close to a quantum phase transition by measuring the so-called excess conductivity $\sigma_H$ in the THz range, being $\sigma_H$ determined by subtracting the BCS contribution $\sigma_1^{\textmd{BCS}}$ from the measured real part $\sigma_1^{\textmd{exp}}$, namely $\sigma_{H} = (\sigma_1^{\textmd{exp}} - \sigma_1^{\textmd{BCS}}$) \cite{dresselnature}.
Here, we make use of the dielectric response as a \emph{working horse} to demonstrate the emergence of what we propose as Higgs-like stiffness, hereafter HLS, on the verge of \emph{any} phase transition. In analogy to the canonical massive Higgs mode, we propose that on the verge of phase transitions, fluctuations of the order parameter amplitude give rise to an increased stiffness due to the emergence of low-energy excitation modes \cite{andersonbook}. Although P.W. Anderson has discussed the concept of generalized rigidity in a broad way \cite{andersonbook}, we have focused on the particular case of the vicinity of phase transitions. In our analysis, the electric dipoles stiffness is enhanced, justifying thus the use of the term HLS. This is a similar situation to the enhancement of the spin stiffness close to a quantum phase transition, for instance, for the case of the one-dimensional Ising model under a transverse magnetic field \cite{prbl}. This is because the spin stiffness $\rho_S \simeq \partial^2 E/\partial \varphi^2$ \cite{sandvik}, where $E$ is the energy and $\varphi$ a reference angle, is linked with the quantum Gr\"uneisen parameter proposed recently by us \cite{prbl}. Since an enhancement of the quantum Gr\"uneisen parameter is observed in the vicinity of quantum critical points/quantum phase transitions \cite{prbl}, we can infer that the spin stiffness is also increased, which emulates the ``creation of mass'' in such a regime. Next, we discuss our proposal in terms of the electric dipole stiffness enhancement in the vicinity of the second-order ferroelectric transition for the (TMTTF)$_2$SbF$_6$ salt as a case of study.

\section*{Results and discussions}

\subsection*{The Landau and $\phi^4$ theories for the second-order ferroelectric transition}
\begin{figure}[!t]
\centering
\includegraphics[width=0.4\columnwidth]{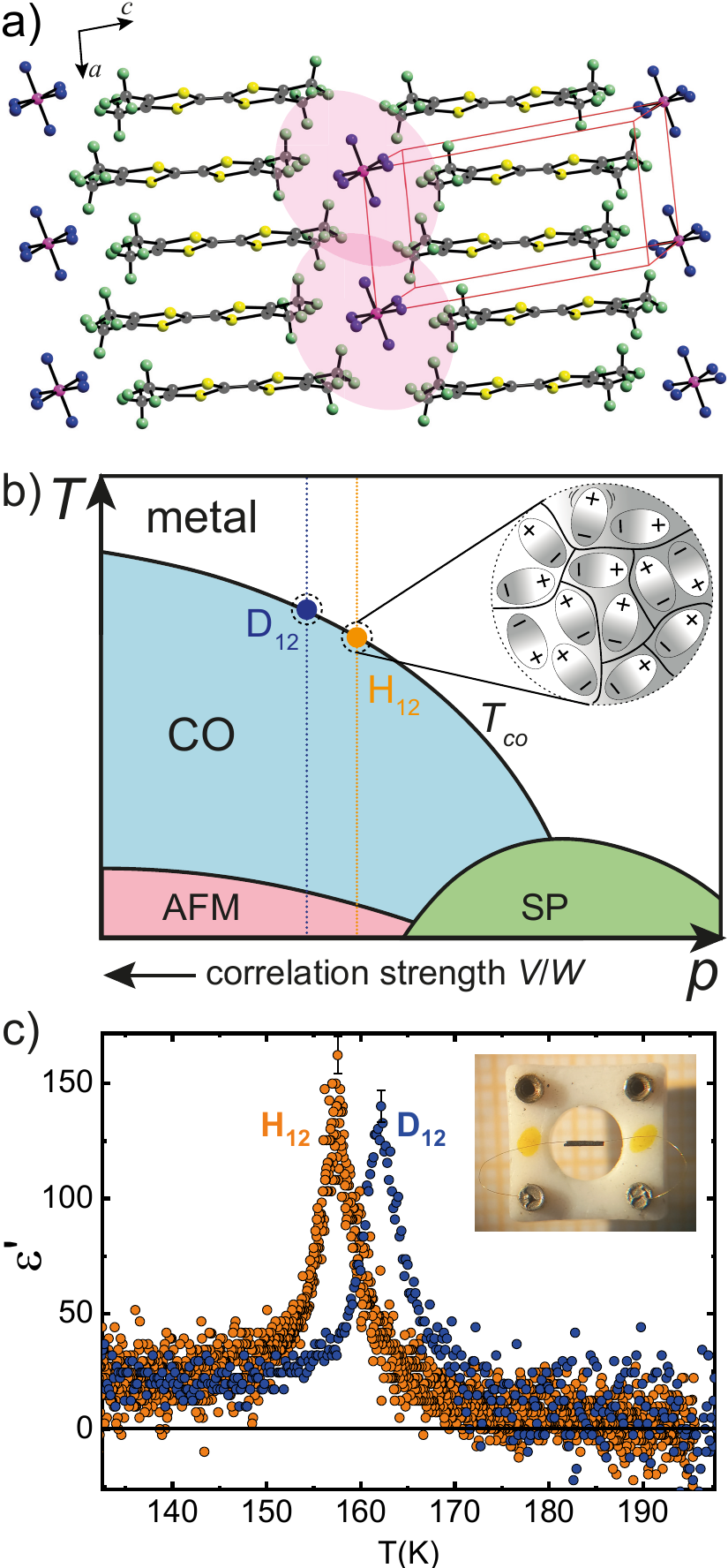}
\caption{\footnotesize a) Crystal structure of (TMTTF)$_2$SbF$_6$, where the spheres represent hydrogen (green), carbon (gray), sulfur (yellow), fluorine (blue), and antimony (purple) atoms. The triclinic unit-cell is outlined by the red solid lines. The cavities delimited by the methyl-end groups are outlined in pink color. b) Schematic temperature $T$ \emph{versus} pressure $p$ phase diagram of the (TMTTF)$_2$SbF$_6$ Fabre salt showing the metallic, charge-ordered (CO), antiferromagnetic (AFM), and spin-Peierls (SP) phases. The vertical dashed lines represent the corresponding charge-ordering transition temperatures $T_{co}$ for both hydrogenated ($H_{12}$) and partially deuterated (D$_{12}$) variants. The correlation strength $V$/$W$ $\propto$ $p^{-1}$ is also depicted, where $V$ is the inter-site Coulomb repulsion and $W$ the bandwidth. The electric dipoles configuration is schematically depicted aiming to indicate fluctuations of the order parameter close to $T_{co}$. Figure based on Ref.\,\cite{pustogow}. c) Quasi-static ($f$ = 1\,kHz) ionic dielectric constant $\varepsilon'$ \emph{versus} $T$ along the $c^*$-axis direction for the hydrogenated (H$_{12}$, orange color circles) and deuterated ($D_{12}$, blue circles) variants of (TMTTF)$_2$SbF$_6$ from raw data set, being measurements taken every $\approx$ 14\,s.  In the inset, a picture of a (TMTTF)$_2$SbF$_6$ sample with gold wires attached on its surface using carbon paste is shown. The gold wires are fixed in an insulating socket with silver paint. Ge varnish is employed to fix the gold wires on the socket to avoid stress on the crystal. Details in the main text.}
\label{Fig-2}
\end{figure}
To showcase our proposal, we start by recalling the phenomenon of dielectric catastrophe for frequencies $\omega \rightarrow 0$ under the light of the Bruggemann effective medium approximation (BEMA) \cite{efros}. Essentially, BEMA approach considers a percolation theory background and describes the behavior of the static dielectric constant close to a phase transition, being recently employed to trace dielectric constant results under pressure close to the Mott transition for the spin-liquid candidate $\kappa$-(BEDT-TTF)$_2$Cu$_2$(CN)$_3$ \cite{rosslhuber,pustogow}. Also, upon crossing the Mott transition by application of pressure, a sign-change of the dielectric constant is observed, being the dielectric response enhanced in the low-frequency regime \cite{pustogow,rademaker}. Motivated by such literature results, our goal here is to make use of the dielectric catastrophe observed at the Mott transition for the spin-liquid candidate as a case of study to demonstrate that upon measuring the real part $\varepsilon'$ of the complex dielectric constant $\hat{\varepsilon} = \varepsilon' + \varepsilon''i$, the HLS can be accessed, being $\varepsilon''$ the imaginary contribution to $\hat{\varepsilon}$. We present $\varepsilon'$ experimental results for the Fabre salt (TMTTF)$_2$SbF$_6$, where TMTTF stands for tetramethyltetrathiafulvalene. The latter presents a metal-to-insulator transition coinciding with the establishment of a Mott-Hubbard ferroelectric phase \cite{reviewmariano,prb2018}, which we consider as an appropriate playground to explore the manifestation of the HLS as a case of study. On the verge of the metal to the Mott-Hubbard ferroelectric phase transition, metallic puddles coexist with an insulating matrix or vice-versa \cite{Mottiscool}.
Let us write the free energy $F$ expansion with pressure $p$ as the tuning parameter considering the electric polarization $P$ as the order parameter in the frame of the well-known Landau theory for phase transitions, namely for $T = T_c$ \cite{lines}:
\begin{equation}
 F = a_0(p-p_c)P^2 + bP^4 + ...,
\label{freeenergy}
\end{equation}
where $a_0$ and $b$ are non-universal constants and $p_c$ the critical pressure, in which the Mott transition takes place \cite{rosslhuber,pustogow}. Note that Eq.\,\ref{freeenergy} is applicable for the spin-liquid candidate where the Mott transition is accessed by application of $p$ \cite{rosslhuber,pustogow}, being that for the (TMTTF)$_2$SbF$_6$ salt the Mott transition is driven by varying  $T$.
For a linear dielectric medium $P = (\hat{\varepsilon}-\varepsilon_0)E$, where $\varepsilon_0$ is the vacuum permittivity and $E$ the electric field  \cite{lines}, being the vector notation omitted for simplicity. In terms of the real and imaginary parts of $\hat{\varepsilon}$ and considering only quadratic terms for a fixed $E$, Eq.\,\ref{freeenergy} becomes:
\begin{equation}
F \simeq a_0(p-p_c)({\varepsilon'}^2 - {\varepsilon''}^2)E^2 + b({\varepsilon'}^2 - {\varepsilon''}^2)^2E^4.
\label{freeenergy2}
\end{equation}
Following discussions in Ref.\,\cite{dresselnature}, we define an excess dielectric constant ${\varepsilon_H}^2 = (\varepsilon'^2 - \varepsilon''^2)$ in analogy with the excess conductivity $\sigma_H$, which is pivotal to understand the manifestation of the HLS. The latter has reminiscence with the so-called excess function, usually employed in Thermodynamics \cite{atkins}. Note that ${\varepsilon_H}^2$ emerges naturally from the consideration of only quadratic terms upon computing $F$. It is remarkable that Eq.\,\ref{freeenergy2} has a similar mathematical structure of the two component potential energy $U$ embedded in the Lagrangian employed in classical field theory to describe continuous symmetry breaking, the so-called $\phi^4$ theory \cite{blundell}:
\begin{equation}
U(\phi_1,\phi_2) = -\frac{\mu^2}{2}({\phi_1}^2 + {\phi_2}^2) + \frac{\lambda}{4!}({\phi_1}^2 + {\phi_2}^2)^2,
\label{potential}
\end{equation}
where $\phi_1$ and $\phi_2$ are the so-called scalar fields, being $\phi_1$ associated with the massive excitation and $\phi_2$ with the massless one, $\mu$ is the mass, and $\lambda$ the interaction strength. Note that $U(\phi_1, \phi_2)$ gives rise to the well-known ``Mexican hat'' shape, see Fig.\,\ref{Fig-1} b). %We are aware of Ginzburg-Landau-type theories for coupled order parameters, cf.\,Ref.\,\cite{suhl}. However, since the character of both metal-to-insulator and ferroelectric transitions is distinct, the two transitions are weakly coupled, so that the $\phi^4$-theory can be employed for the second-order ferroelectric transition.}
In our analysis, by comparing Eqs.\,\ref{freeenergy2} and \ref{potential}, we associate $\varepsilon'$ with $\phi_1$. It is to be noted that depending on the convention one can define $\hat{\varepsilon} = (\varepsilon' - \varepsilon''i)$ \cite{varadan}, so that $F \simeq a_0(p-p_c)({\varepsilon'}^2 + {\varepsilon''}^2) + b({\varepsilon'}^2 + {\varepsilon''}^2)^2$. Hence, we propose at this point a connection between the dielectric catastrophe and the natural manifestation of the HLS, i.e., pseudo ``creation of mass'', when a continuous symmetry breaking takes place accompanying a particular phase transition. It is worth mentioning that we have measured the dielectric response $\varepsilon'$ associated with the ionic contribution, i.e., along the $c^*$-axis direction, of the second-order ferroelectric transition \cite{nadmonceau,Monceau2}. This is a very important aspect since the ionic contribution can be treated as weakly coupled to the charge localization/disproportionation along the stacks formed by the TMTTF molecules. Thus, we apply the $\phi^4$ theory solely for the ferroelectric transition.

\subsection*{Connection between the HLS and the dielectric response}
Figure\,\ref{Fig-2} c) depicts experimental results of the ionic contribution to $\varepsilon'$ normalized by $\varepsilon_0$ \emph{versus} $T$ for the hydrogenated and unprecedentedly for the partially deuterated (97.5\%) variants of the molecular metal (TMTTF)$_2$SbF$_6$. Such data set depicted in Fig.\,\ref{Fig-2} c) aims to provide insights about the dielectric response $\varepsilon'$ for a real system. %, which enables us to employ two complementary theoretical approaches to sustain our proposal of HLS.} %This particular salt of the TMTTF series was chosen because the metal-to-insulator transition coincides with the charge-ordering transition \cite{reviewmariano}.
In a previous work, we have reported on the ionic contribution of $\varepsilon'$ to the ferroelectric phase for the hydrogenated variant of (TMTTF)$_2$SbF$_6$, among other Fabre salts \cite{prb2018}. %\textcolor{blue}{The capacitance $C$ of the measured single-crystals was measured along the $c^*$-axis using the two-points method. Upon measuring $C$, the dielectric constant $\varepsilon$ was determined by employing the well-known relation from textbooks for a parallel plates capacitor $\varepsilon = Cd/A$, where $d$ is the distance between the capacitor plates and $A$ the the capacitor plates area, being in our case, respectively, the thickness and the area of the investigated salt. Finally, we have computed $\varepsilon' = \varepsilon/\varepsilon_0$.}
Clear peak-like anomalies are observed at the charge-ordering transition temperature $T_{co}$ = 157 and 162\,K, respectively, for hydrogenated and partially deuterated variants \cite{nadmonceau}, being a similar shift in $T_{co}$ due to deuteration effects also observed for the case of the (TMTTF)$_2$PF$_6$ system \cite{prb2018}. Based on our analyzis in the frame of the $\phi^4$ theory, we propose that the peak-like anomaly in $\varepsilon'$ at $T_{co}$ is related to the emergence of an enhanced electric dipole stiffness into the system, i.e., ``creation of mass''. Although the (TMTTF)$_2$SbF$_6$ salt is the one that presents a behavior closer to a mean field-type (MF) transition, it deviates from MF most likely due to the presence of impurities, dislocations, and structural defects in the specimen \cite{nadmonceau}. Also, the HLS can be suppressed by X-ray irradiation since it gradually annihilates the dielectric response \cite{prb2018}. Yet, the Curie law does not account for negative $\varepsilon'$, which is observed experimentally in the vicinity of $T_{co}$, cf.\,Fig.\,\ref{Fig-2} c). The order-disorder character is intrinsic to the coexistence region and thus the corresponding fluctuations of the order parameter in the vicinity of $T_{co}$. Following Ref.\,\cite{crystalspouget}, each counter-anion lies in a double-well potential and the minima are linked with the formation of hydrogen bonds between the counter-anion and the methyl-end groups. Deuteration acts like a negative pressure making the cavity delimited by the methyl-end groups bigger \cite{chipouget}, increasing thus the $V$/$t$ ratio, where $V$ is the inter-site Coulomb repulsion and $t$ the transfer-integral between nearest neighbor sites \cite{nadmonceau}. The potential height is increased upon deuteration and thus $T_{co}$ is enhanced for the deuterated variant \cite{chipouget,prb2018} [Fig.\,\ref{Fig-2} b)], %The magnitude of the peak-like anomaly in $\varepsilon'$ for the deuterated variant is somewhat lower ($\thicksim$ 20\,\%).
%Such a feature is directly related to the increased mass upon deuteration of the methyl-end groups of the TMTTF molecule,
making the system to reduce its dynamics naturally and thus its dielectric response is expected to be less pronounced when compared with the hydrogenated variant for a given frequency. A similar behavior was also observed by Nad and collaborators regarding the electronic dielectric response contribution at the Mott-Hubbard ferroelectric phase transition \cite{nadmonceau}, which is about three orders of magnitude higher than the ionic one \cite{prb2018}. Also, Nad demonstrated experimentally that for the TMTTF salts the dielectric response is dramatically enhanced upon reducing the frequency of the applied electric field \cite{nadmonceau}, a feature also observed for other molecular salts, such as the quasi-two-dimensional system $\kappa$-(BEDT-TTF)$_2$Cu[N(CN)$_2$]Cl \cite{naturemariano}. Note that for  $T > T_{co}$, negative values of $\varepsilon'$ are attained due to the metallic character of the system in this temperature range \cite{dresselbook}. We assign the enhanced data dispersion in this $T$-range to the intrinsic fluctuations of the order parameter in this regime.  It is clear that the maximum in $\varepsilon'$ is associated with the canonical Landau free energy variation $\Delta F = -EP + g_0 + 1/2g_2P^2 + ...$ close to $T_c$, being $g_0$ and $g_2$ the Landau coefficients, cf.\,\cite{andersonbook} and references cited therein. Since $\varepsilon'$ is related to $P$, it becomes also clear that $\Delta F$ scales with $\varepsilon'$, being maximized at $T_{co}$ and vanishing above and below $T_{co}$. In other words, fluctuations of the order parameter are mainly governed by $\varepsilon'$, cf.\,black thick line depicted in Fig.\,\ref{Fig-1} b), while we associate the massless mode with $\varepsilon''$ indicated by the cyan line in Fig.\,\ref{Fig-1} b). Considering that the dielectric response is dominated by $\varepsilon'$, i.e., $\varepsilon_H^2 \simeq \varepsilon'^2$ for $\varepsilon'' \ll \varepsilon'$, the Mexican hat shape is not altered. At this point, a natural question is: why do we have considered $\varepsilon''$ not dominant in the present analysis? This is because $\varepsilon''$ is associated with the dissipative component of $\hat{\varepsilon}$, so that we assume that fluctuations of the order parameter are slightly affected by $\varepsilon''$. If the dissipative component dominates the system, no finite electric polarization would be observed \cite{dresselbook}. It is clear that for any complex physical quantity, the imaginary contribution is always finite. Yet, since the dielectric constant and the refractive index are related quantities, the $\varepsilon'$ peak-like anomaly at $T_{co}$ indicates the onset of critical opalescence across the Mott transition \cite{opalescence}. Now, we make an analyzis considering the well-known complex order parameter of the form $\Psi=Ae^{i\theta}$, being the amplitude $A = ({\phi_1}^2+{\phi_2}^2)^{1/2}$ and the phase $\theta=\textmd{arctan}(\phi_2/\phi_1)$ \cite{philip}. Just to mention, the latter is reminiscent of the Weinberg angle $\theta_W$ in the frame of the electroweak theory \cite{piers}. Based on our proposal of HLS, we can analogously write $\Psi= ({\varepsilon'}^2+{\varepsilon''}^2)^{1/2}e^{i\, [\textmd{arctan}(\varepsilon''/\varepsilon')]}$, so that it becomes evident that a finite $\varepsilon''$ affects both the amplitude and phase of the complex order parameter. At this point, it is appropriated to recall that for the case of superconductors and superfluids, the phase $\phi$ bending energy $U_{b} \propto (\nabla\phi)^2$ \cite{piers}. In our case, the HLS can be thus quantified in terms of $[\nabla\textmd{arctan}(\varepsilon''/\varepsilon')]^2$. Note, however, that the behavior of the imaginary contribution to the response function is system-dependent and not all response functions give rise to a finite imaginary part \cite{pippard}. The latter is finite only if it is coupled to the microscopic quantity associated with the corresponding ordering \cite{pippard}. Following discussions by P. Coleman \cite{piers}, gauge field + phase $\rightarrow$ massive gauge field. In our proposal of HLS,  the local electric field that emerges due to the inherent interactions between electric dipoles in the vicinity of the phase transition plays the role of the gauge field and the phase is $\theta = \textmd{arctan}(\varepsilon''/\varepsilon')$, giving rise to the ``creation of mass'' in terms of HLS. Aiming to illustrate our HLS proposal, following discussions in Ref.\,\cite{pippard}, we consider the simple case of a particle with mass $m$ under a potential $U(x) = -ax^2 + cx^4$, where $a$ and $c$ are arbitrary constants. For small oscillations around the equilibrium position, the frequency $\omega^2 = 2|a|/m$ ($a < 0$) and $\omega^2 = 4a/m$ ($a > 0$) \cite{pippard}. For $a \rightarrow 0$, the period of oscillation $T^* \propto |a|^{-1/2} \rightarrow \infty$, which implies a critical slowing down \cite{pippard} that emulates our proposal of HLS as if there was a ``creation of mass''. %\textcolor{red}{\st{Furthermore, based on the fact that the optical conductivity $\sigma = \varepsilon\omega\varepsilon_0$, a link between the real part of $\hat{\sigma}$ and $\hat{\varepsilon}$ can be made. Hence, we are able to make a direct connection with the findings}}

The authors of Ref.\,\cite{dresselnature} probed the Higgs mode in an insulator to superconductor quantum phase transition. In the latter, thin films of NbN and InO were tuned close to quantum criticality upon increasing disorder. Note that for the case of the (TMTTF)$_2$SbF$_6$ salt, the emergence of metallic puddles in an insulating matrix, or vice-versa, i.e., in the phases coexistence region, is analogous to the insertion of disorder. We propose that \emph{any} complex physical quantity can be employed to probe the HLS on the verge of phase transitions, e.g., refractive index and a.c.\,magnetic susceptibility, being the choice of the observable dependent on the particular investigated system. In other words, a similar analysis based on Eq.\,\ref{freeenergy2} and above discussions can be performed for \emph{any} complex physical quantity by considering its relation with the order parameter of the particular investigated phase transition. It is clear that the form of Eq.\,\ref{freeenergy2} remains the same for any complex physical quantity. Our proposal can be also extended to complex thermodynamical quantities, such as complex specific heat \cite{mario,alnot}. Regarding the universality of our approach, note that the role played by the HLS is distinct depending on the particular phase transition considered, for instance, in the ferroelectric case, HLS is linked with the electric dipole stiffness, while for a magnetic phase transition, the spin stiffness plays the role of HLS. The corresponding excess function for a given phase transition should be defined accordingly.

\subsection*{The HLS under the perspective of the Mott transition}
Aiming to support the discussions regarding the HLS in the frame of the $\phi^4$ theory, we briefly discuss in the following a few general aspects related to the Higgs mechanism in superconductors and analyze the HLS using the plasma frequency and Drude weight.  Interestingly, time-dependent Ginzburg-Landau theory for superconductivity delivers the relaxation time $\tau \propto (T - T_c)^{-1}$ \cite{tinkham}, so that it becomes clear that for $T \rightarrow T_c \Rightarrow \tau \rightarrow\infty$. Such a behavior is consistent with the emulation of ``creation of mass'' on the verge of a superconducting transition. Following Anderson, close to $T_c$, fluctuations of the order parameter amplitude are enhanced, implying that excitations have become concentrated at low-frequencies \cite{andersonbook}. A universal aspect of phase transitions is that close to $T_c$, given the phases competition in this regime, we are dealing with a phases coexistence region, so that $\tau$ can be computed employing Avramov/Casalini's model \cite{Mottiscool,avramov1}. In such cases, $\tau = \tau_0\exp{(C/Tv^{\Gamma_{eff}})}$, where $\tau_0$ is a typical relaxation time of the system, $C$ a non-universal constant, $v$ the volume, and $\Gamma_{eff}$ the effective Gr\"uneisen parameter \cite{Mottiscool,avramov1}. Regarding the Mott transition, the dielectric response $\varepsilon'$ as a function of pressure is positive (negative) when the system is a Mott insulator (metal) \cite{rosslhuber,pustogow}. Recently, some of us have demonstrated that upon crossing the first-order transition line of the Mott transition, $\tau$ is dramatically enhanced and an electronic Griffiths-like phase sets in \cite{Mottiscool}, so that ``mass creation'' can also be inferred. It is well-known from textbooks that the electron mass $m_e$ is related to the plasma frequency $\omega_p$ by $m_e = e^2 n_e/\varepsilon_0{\omega_p}^2$ \cite{dresselbook}, where $e$ is the electron fundamental charge, and $n_e$ the electron density. It turns out that upon tuning the system from a metal to a Mott insulator, electrons tend to localize suppressing thus charge fluctuations, i.e., $\omega_p \rightarrow 0$ $\Rightarrow$ $m_e \rightarrow \infty$. The latter is corroborated by recent random phase approximation (RPA) calculations of charge collective modes at the onset of the Mott transition \cite{fresard} and it is in line with our proposal of the emergence of the HLS, i.e., ``mass enhancement'', on the verge of a phase transition, cf.\,proposed by Anderson \cite{Anderson1963}. In terms of the Drude weight $D =\lim_{\omega\rightarrow0}{\omega\sigma_2(\omega)} \simeq (\partial^2 E_0/\partial k^2) = \hbar^2m^{-1}(k)$, where $E_0$ the ground-state energy, $k$ the wave vector, and $m(k)$ the effective mass \cite{kohn,resta}. Upon going from a metal to an insulating phase, $(\partial^2 E_0/\partial k^2) \rightarrow 0 \Rightarrow D \rightarrow 0$, $(\partial^2 E_0/\partial k^2) \rightarrow 0 \Rightarrow m(k) \rightarrow \infty$. The latter is reminiscent of the breakdown of the Hellmann-Feynman theorem on the verge of a quantum phase transition \cite{prbl}. %Also, upon a continuous pressurization, a phase coexistence region embodying Mott insulating and metallic puddles sets in \cite{rosslhuber,pustogow, Mottiscool}. It is thus natural that upon crossing the first-order transition line within the phases coexistence region, going from the insulating to the metallic dome, $\varepsilon'$ changes sign from positive to negative, cf.\,experimental results depicted in Fig.\,\ref{Fig-2} c).

\subsection*{About fractonic excitations and the HLS}\label{fractons}
Now, we discuss the ``mass enhancement'', i.e., the slow-dynamics regime, in connection with fractons, which are linked with immobile excitations \cite{hermele}. In our approach, such fractonic excitations can be connected with the enhancement of the electric dipole stiffness \cite{kohn,hughes}, so that the appearance of the HLS might be linked with the increased immobility of the electric dipoles on the verge of the Mott transition. The enhancement of the dipole stiffness particularly in the vicinity of a quantum critical point was already reported \cite{hughes}, but, to the best of our knowledge, no association with the concept of Higgs mode was discussed so far. For the Fabre salts, the methyl-end groups play a crucial role in the stabilization of the Mott-Hubbard ferroelectric phase \cite{reviewmariano,shunsuke,reviewpouget2}. Regarding the particular case of the (TMTTF)$_2$SbF$_6$, the charge disproportionation 2$\delta$ = 0.5 \cite{zamborszky}, so that the ferroelectric phase for the SbF$_6$ salt is one the most robust among the Fabre salts.
%This is because the high polarizability/electronegativity of Sb induces a more pronounced charge disproportionation, making the dielectric response of (TMTTF)$_2$SbF$_6$ to be enhanced when compared to the other salts \cite{prb2018}.
%Such a ferroelectric phase is most likely accompanied by a partial freezing/immobility of the so-called rigid unit modes (RUM) \cite{goodwin}, which are associated with rotational degrees of freedom of the counter-anions \cite{reviewmariano}. Given the crystallographic peculiarity of the $c^*$-axis, along which layers of donor molecules are separated by layers of counter-anions, its thermal expansivity is more pronounced \cite{reviewmariano}. It is to be noted that along the $c^*$-axis we are probing the ionic dielectric response \cite{prb2018}. The counter-anion displacement accompanying the Mott-Hubbard ferroelectric transition gives rise to an spontaneous symmetry breaking. Following discussions in Ref.\,\cite{shunsuke}, there is a huge charge accumulation on the methyl-end groups. Hence, the presence of various degrees of freedom in the here-discussed molecular salts and the rich phase diagram enables the exploration of many exotic manifestations of matter in condensed matter Physics.}
Based on the fact that the Mott-Hubbard charge disproportionation occurs along the TMTTF stacks \cite{shunsukecrystals}, it is natural to infer that the counter-anion displacement from its center-symmetric position gives rise to the appearance of an ionic contribution to the dielectric response \cite{prb2018}. The latter is most likely related to the formation of a charge imbalance between the methyl-end groups and the counter-anion. This scenario is corroborated by vibrational spectra measurements showing that a distortion of the octahedral structure of the counter-anions takes place below $T_{co}$ \cite{dresselcrystals}. More specifically, the low structural symmetry, i.e., triclinic  P$\bar{\textmd{1}}$ space of the Fabre salts produces an intricate counter-anion displacement within the cavities formed by the methyl-end groups, cf.\,Fig.\,\ref{Fig-2} a).
%It is to be noted that the key role played by counter-anions is corroborated by fluorine 1s core level spectra of (TMTTF)$_2$SbF$_6$ \cite{medjanik}.
Following discussions in Ref.\,\cite{reviewmariano}, there are two possibilities of rearrangement of the counter-anions, i.e., the counter-anion displacement might occur approaching either methyl-end groups or the sulfur atom of the TMTTF molecule nearby. Assuming that the counter-anion displacement occurs towards the methyl-end groups, such entities will build up an ``ionic'' electric dipole. In this context, both counter-anion and methyl-end groups become rigid, i.e., their degree of immobility is enhanced, emulating thus the manifestation of fractonic excitations \cite{hermele}. This is corroborated by the experimental verification of a divergent relaxation time at $T_{co}$ for the (TMTTF)$_2$AsF$_6$ salt \cite{asf6slowingdown}. It is also to be noted that the freezing of the methyl-groups rotational degrees of freedom around $T_{co}$ can be inferred from high-resolution thermal expansion \cite{reviewmariano} and $^{19}$F NMR measurements \cite{zamborszky,nakamura}. Regarding thermal expansion measurements for the Fabre salts, the continuous freezing of the rotational degrees of freedom of the methyl-end groups makes such vibrational modes no longer to contribute to the thermal expansion \cite{reviewmariano}. In this regime, the conservation of the electric dipole moment restricts the rotational degrees of freedom of both counter-anion and methyl-end groups, which we interpret as the emergence of fractons \cite{hermele}. It is worth mentioning that the inherent competition between various degrees of freedom in molecular conductors enables us to explore the here-discussed fascinating manifestations of matter.

\section*{Conclusions}
In summary, we have demonstrated that close to phase transitions the HLS appears naturally as a consequence of strong fluctuations of the order parameter. We propose that the deviation of $\varepsilon'$ from its ideal behavior due to a finite $\varepsilon''$ emulates the electric dipole stiffness in our case of study, being extendable to any phase transition with its corresponding complex observable. Experimental results for the (TMTTF)$_2$SbF$_6$ salt suggest that the quasi-static ionic dielectric response is linked with an increased immobility of both the counter-anion and the methyl-end groups, which we associate with the emergence of fractonic excitations. %We propose that \emph{any} complex physical quantity can be employed to explore the HLS in the vicinity of phase transitions. We showcase our proposal based on two complementary lines of action, one associated with the $\phi^4$ theory and another with respect to the plasma frequency.
Our proposal of HLS is universal and it is challenging to explore this intricate manifestation of matter on the verge of other phase transitions, being highly desirable to determine the gauge field responsible for the emergence of HLS in each case.

\section*{Methods}

The electrical contacts were prepared using tempered gold wires provided by Cryogenics with 20\,$\mu$m diameter properly attached to the specimen surfaces along the $c^*$-axis direction employing carbon paste. The sample dimensions are (2.3 $\times$ 0.6 $\times$ 0.3)\,mm$^3$ and (2.5 $\times$ 0.4 $\times$ 0.1)\,mm$^3$, respectively, for the specimens of hydrogenated and partially deuterated (TMTTF)$_2$SbF$_6$. The measurements were carried out employing a cooling rate of $-$10\,K/h (hydrogenated) and $-$40\,K/h (deuterated), an electric field of 500\,mV/cm at a fixed frequency of 1\,kHz. A high-resolution Andeen-Hagerling capacitance bridge and a Teslatron PT cryostat supplied by Oxford Instruments were used. For further experimental details, please refer to \cite{prb2018}. The figures were generated employing the softwares Wolfram Mathematica version 13.0 (License ID L8671-6484), Origin version 2024b (Registration ID SWX-9OF-FFY), and Adobe Illustrator 2024 version 28.5.

%\vspace{0.5cm}

\section*{Acknowledgements}
MdeS acknowledges financial support from the S\~ao Paulo Research Foundation – Fapesp (Grants number 2011/22050-4, 2017/07845-7, and 2019/24696-0), National Council of Technological and Scientific Development – CNPq (Grants number 303772/2023-9), J.P. Pouget and Pascale Foury-Leylekian for discussions, and Alec Moradpour (in memorian) for synthesizing the investigated single-crystals. ACS acknowledges CNPq (Grants number 308695/2021-6). LS acknowledges IGCE for the post-doc fellowship.

\section*{Author contributions}
L.S. and M.deS. carried out the experiments and calculations and L.S. generated the figures. L.S. and M.deS. wrote the paper with contributions from A.C.S. and R.E.L.M. All authors revised the manuscript. M.deS. conceived and supervised the project.

\section*{Competing interests}
The authors declare no competing interests.

\section*{Data availability}
The datasets generated during and/or analysed during the current study are available from the corresponding author on reasonable request.

\end{document}